\documentstyle[aps,epsfig,floats]{revtex}

\tighten
\newcommand{\be}{\begin{equation}}
\newcommand{\ee}{\end{equation}}
\newcommand{\bea}{\begin{eqnarray}}
\newcommand{\eea}{\end{eqnarray}}

\newcommand{\bm}[1]{\bbox{#1}}

\def\st{{\scriptscriptstyle T}}
\def\slash{\rlap{/}}

\begin{document}

\title{
\begin{flushright}
\begin{minipage}{4 cm}
\small
VUTH 99-26
\end{minipage}
\end{flushright}
Bounds on transverse momentum dependent distribution\\ and fragmentation
functions.}

\author{A. Bacchetta, M. Boglione, A. Henneman and P.J. Mulders}
\address{\mbox{}\\
Department of Theoretical Physics, Faculty of Science, Vrije Universiteit\\
De Boelelaan 1081, NL-1081 HV Amsterdam, the Netherlands
}

\maketitle

\begin{abstract}
We give bounds on the distribution and fragmentation functions that
appear at leading order in deep inelastic 1-particle inclusive
leptoproduction or in Drell-Yan processes. These bounds simply follow
from positivity of the defining matrix elements and are an important
guidance in estimating the magnitude of the azimuthal
and spin asymmetries in these processes.
\end{abstract}


In deep-inelastic processes the transition from hadrons to 
quarks and gluons is described in terms of distribution and
fragmentation functions. For instance, at leading order in the 
inverse hard scale $1/Q$, the cross section for inclusive 
electroproduction $e^- H \rightarrow e^- X$
is given as a charge squared weighted sum over quark
distribution functions, which describe the probability of finding quarks
in hadron $H$. In electron-positron annihilation, the 1-particle inclusive
cross section for $e^+e^- \rightarrow hX$ is given as a charge squared
weighted sum over quark and antiquark fragmentation functions,
describing the decay of the produced (anti)quarks into hadron $h$.

\begin{figure}[t]
\begin{minipage}{7.5 cm}
\begin{center}
\epsfig{file=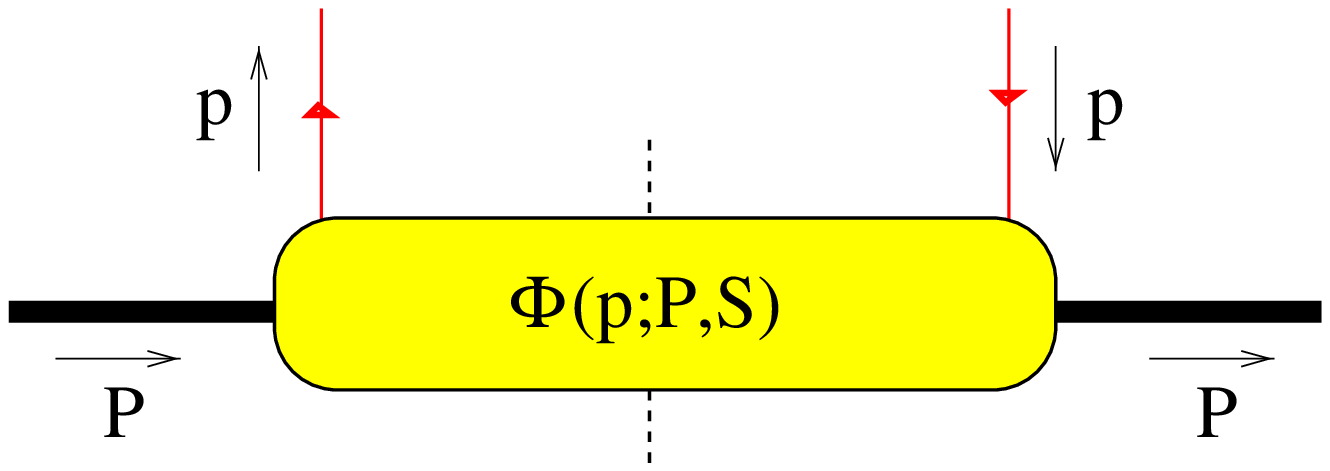,width = 5.5 cm}
\end{center}
\caption{\label{fig1}
Matrix element for distribution functions.}
\end{minipage}
\hspace{0.5cm}
\begin{minipage}{7.5 cm}
\begin{center}
\epsfig{file=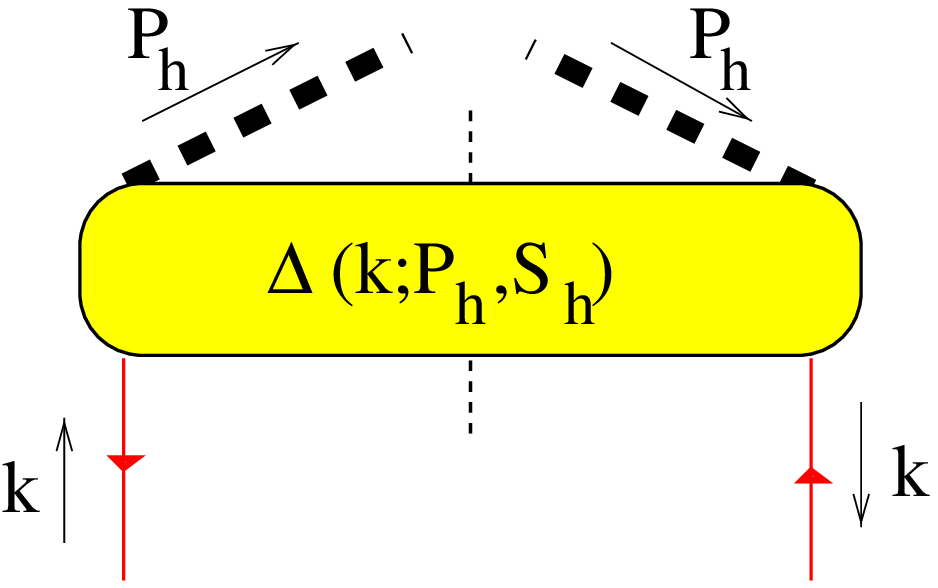,width = 3.5 cm}
\end{center}
\caption{\label{fig2}
Matrix element for fragmentation functions.}
\end{minipage}
\end{figure}

The distribution functions for a quark can be
obtained from the lightcone correlation 
function~\cite{Soper77,Jaffe83,Manohar90}
\be
\Phi_{ij}(x) = \left. \int \frac{d\xi^-}{2\pi}\ e^{ip\cdot \xi}
\,\langle P,S\vert \overline \psi_j(0) \psi_i(\xi)
\vert P,S\rangle \right|_{\xi^+ = \xi_\st = 0},
\ee
depending on the lightcone fraction $x = p^+/P^+$.
To be precise, the lightlike directions $n_+$ and $n_-$,
satisfying $n_+^2 = n_-^2$ = 0 and $n_+\cdot n_-$ = 1, define
the lightcone coordinates $a^\pm = a\cdot n_\mp$. The hadron
momentum $P$ is chosen so that it has no components orthogonal
to $n_+$ or $n_-$, thus the transverse hadron momentum $P_\st = 0$.
The correlator contains the soft parts appearing in hard scattering
processes, and is related to the forward amplitude for
antiquark-hadron scattering (see Fig.~\ref{fig1}).
The relevant part is $\Phi\slash n_-$ = $\Phi\gamma^+$. 
Inserting a complete set of intermediate states and generalizing to
off-diagonal spin, one obtains
\bea
(\Phi\gamma^+)_{ij,s^\prime s}
& = & \left. \int \frac{d\xi^-}{2\pi\sqrt{2}}\ e^{ip\cdot \xi}
\,\langle P,s^\prime\vert \psi^\dagger_{+j}(0) \psi_{+i}(\xi)
\vert P,s\rangle \right|_{\xi^+ = \xi_\st = 0}
\nonumber \\
& = & \frac{1}{\sqrt{2}}\sum_n
\langle P_n\vert \psi_{+j}(0)\vert P,s^\prime\rangle^\ast
\langle P_n\vert \psi_{+i}(0)\vert P,s\rangle
\,\delta\left(P_n^+ - (1-x)P^+\right) ,
\label{dens}
\eea
where $\psi_+ \equiv P_+\psi = \frac{1}{2}\gamma^-\gamma^+\psi$ is the
good component of the quark field~\cite{KS70}. This representation
shows that the correlation functions have a natural interpretation
as lightcone momentum densities. 

In order to study the correlation function in a spin 1/2 target one 
introduces a spin vector $S$
that parametrizes the spin density matrix $\rho (P,S)$. It satisfies
$P\cdot S$ = 0 and $S^2 = -1$ (spacelike) for a pure state, $-1 <
S^2 \le 0$ for a mixed state. Using $\lambda \equiv MS^+/P^+$ and 
the transverse spin vector $S_\st$, the condition becomes 
$\lambda^2 + \bm S_\st^2 \le 1$, as can be seen from the rest-frame
expression $S$ = $(0,\bm S_\st, \lambda)$. The precise equivalence
of a $2\times 2$ matrix
$\tilde M_{ss^\prime}$ in the target spin space
and the $S$-dependent function $M(S)$ is
$M(S) = \mbox{Tr}\left[ \rho(S)\,\tilde M\right]$.
Explicitly, the $S$-dependent function
$M(S)$ = $M_O + \lambda\,M_L + S_\st^1\,M_\st^1
+ S_\st^2\,M_\st^2$,
corresponds to a matrix, which
in the target rest-frame with as basis the spin 1/2 
states with $\lambda = +1$ and $\lambda = -1$ becomes
\be
\tilde M_{ss^\prime} =
\left\lgroup \begin{array}{cc}
M_O + M_L & M_\st^1 - i\,M_\st^2 \\
& \\
M_\st^1 + i\,M_\st^2 & M_O - M_L \\
\end{array}\right\rgroup
\label{spinexplicit}
\ee
From Eq.~\ref{dens} follows that after transposing in Dirac space,
and subsequently extending the matrix $M(S)$ = $(\Phi\gamma^+)^T$ 
to the target spin space gives a matrix in the combined
Dirac $\otimes$ target spin space
which satisfies $v^\dagger M v \ge 0$ for any vector $v$ in that combined
space.

The most general form for the quantity $\Phi \gamma^+$ for a spin 1/2
target in terms of the spin vector is
\be
\Phi(x) \gamma^+ = \Bigl\{
f_1(x) + \lambda\,g_1(x)\,\gamma_5  + h_1(x)\,\gamma_5\,\slash S_\st
\Bigr\}P_+ ,
\label{phiint}
\ee
where the functions $f_1$, $g_1$ and $h_1$ are the leading order
quark distribution functions~\cite{remark}. 
By tracing over the Dirac indices one
projects out $f_1$, which is the quark momentum density 
(see Eq.~\ref{dens}).
By writing $\gamma_5$ as the 
difference of the chirality projectors 
$P_{R/L}$ = $\frac{1}{2}(1\pm \gamma_5)$ it follows that in a 
longitudinally polarized target ($\lambda \ne 0$) $g_1$ 
is the difference of densities for right-handed and left-handed quarks.
By writing $\gamma^i \gamma_5$ as the difference of
the transverse spin projectors $P_{\uparrow/\downarrow}$
= $\frac{1}{2}(1\pm \gamma^i\gamma_5)$, it follows that in a transversely
polarized target ($S_\st \ne 0$) $h_1$ is the difference of quarks with 
tranverse spin along and opposite the target 
spin~\cite{Artru,Cortes92,JJ92}. 
Since $f_1(x)$ is the sum of the densities it is positive and gives
bounds $\vert g_1(x)\vert \le f_1(x)$ and $\vert h_1(x)\vert \le f_1(x)$.

By considering the combined Dirac $\otimes$ target spin space stricter
bounds can be found. As mentioned above, we need to consider the
function $M(S)$ = $(\Phi\,\gamma^+)^T$ in Dirac space.
For this we use a chiral representation. In that representation
the good projector $P_+$ only leaves two (independent) dirac spinors, one
right-handed (R), one left-handed (L).
On this (2-dimensional) basis of good R and L spinors
the matrix $M = (\Phi(x)\,\gamma^+)^T$ obtained from
Eq.~\ref{phiint} is given by
\be
M_{ij} =
\left\lgroup \begin{array}{cc}
f_1(x) + \lambda\,g_1(x) &  (S_\st^1+i\,S_\st^2)\,h_1(x) \\
& \\
(S_\st^1-i\,S_\st^2)\,h_1(x) & f_1(x) - \lambda\,g_1(x)
\end{array}\right\rgroup
\ee
Next we make the spin-structure of the target explicit as
outlined in Eq.~\ref{spinexplicit}, yielding
on the basis $+R$, $-R$, $+L$ and $-L$
\be
\tilde M =
\left\lgroup \begin{array}{cccc}
f_1 + g_1 & 0 & 0 & 2\,h_1 \\
& \\
0 & f_1 - g_1 & 0 & 0 \\
& \\
0 & 0 & f_1 - g_1 & 0 \\
& \\
2\,h_1 & 0 & 0 & f_1 + g_1
\end{array}\right\rgroup .
\ee
From the
positivity of the diagonal elements one recovers the trivial bounds
$f_1(x) \ge 0$ and $\vert g_1(x) \vert \le f_1(x)$, but
requiring the eigenvalues of the matrix to be positive gives
the stricter Soffer bound~\cite{Soffer73},
\be
\vert h_1(x)\vert \le \frac{1}{2}\left( f_1(x) + g_1(x)\right) .
\ee

Analogously bounds can be obtained for transverse momentum 
dependent distribution and fragmentation functions.
Transverse momenta of partons play an important role in hard processes 
with more than one hadron~\cite{RS79}. Examples are 
1-particle inclusive deep-inelastic electroproduction,
$e^- H \rightarrow e^- h X$~\cite{MT96}, or
Drell-Yan scattering, $H_1 H_2 \rightarrow \mu^+\mu^- X$~\cite{TM95}.

The soft parts involving the distribution functions are
contained in the lightfront correlation function
\be
\Phi_{ij}(x,\bm p_T) =
\left. \int \frac{d\xi^-d^2\bm \xi_T}{(2\pi)^3}\ e^{ip\cdot \xi}
\,\langle P,S\vert \overline \psi_j(0) \psi_i(\xi)
\vert P,S\rangle \right|_{\xi^+ = 0},
\ee
depending on $x=p^+/P^+$ and the quark transverse 
momentum $\bm p_\st$ in a target with $P_\st = 0$. 
For the description of quark fragmentation one needs~\cite{CS82}
\be
\Delta_{ij}(z,\bm k_\st) =
\left. \sum_X \int \frac{d\xi^-d^2\bm \xi_\st}{(2\pi)^3} \,
e^{ik\cdot \xi} \,Tr  \langle 0 \vert \psi_i (\xi) \vert P_h,X\rangle
\langle P_h,X\vert\overline \psi_j(0) \vert 0 \rangle
\right|_{\xi^+ = 0},
\ee
(see Fig.~\ref{fig2}) 
depending on $z = P_h^+/k^+$ and the quark transverse momentum $\bm 
k_\st$ leading to a hadron with $P_{h\st} = 0$. A simple boost shows 
that this is equivalent to a quark producing a hadron with transverse 
momentum $P_{h\perp} = -z\,k_\st$ with respect to the quark.
Notice that the expressions given here are in a
lightcone gauge $A^+ = 0$. In a general gauge, a gauge link running
along $n_-$ needs to be included. In the presence of transverse momentum
dependence~\cite{BM99} and hence separation in $\xi_\st$, the links run to
lightcone infinity $\xi^- = \pm \infty$.

Separating the terms corresponding to unpolarized ($O$), 
longitudinally polarized ($L$) and transversely polarized targets
($T$), the most general parametrizations {\em with $p_\st$-dependence}, 
relevant at leading order, are
\bea
\Phi_O(x,\bm p_\st)\,\gamma^+ & = &
\Biggl\{
f_1(x,\bm p_\st^2)
+ i\,h_1^\perp(x,\bm p_\st^2)\,\frac{\slash p_\st}{M}
\Biggr\} P_+
\\
\Phi_L(x,\bm p_\st)\,\gamma^+ & = &
\Biggl\{
\lambda\,g_{1L}(x,\bm p_\st^2)\,\gamma_5
+ \lambda\,h_{1L}^\perp(x,\bm p_\st^2)
\gamma_5\,\frac{\slash p_\st}{M}
\Biggr\} P_+
\\
\Phi_T(x,\bm p_\st)\,\gamma^+  & = &
\Biggl\{
f_{1T}^\perp(x,\bm p_\st^2)\,\frac{\epsilon_{\st\,\rho \sigma}
p_\st^\rho S_\st^\sigma}{M}
+ g_{1T}(x,\bm p_\st^2)\,\frac{\bm p_\st\cdot\bm S_\st}{M}
\,\gamma_5
\nonumber \\ & &\mbox{}
+ h_{1T}(x,\bm p_\st^2)\,\gamma_5\,\slash S_\st
+ h_{1T}^\perp(x,\bm p_\st^2)\,\frac{\bm p_\st\cdot\bm S_\st}{M}
\,\frac{\gamma_5\,\slash p_\st}{M}
\Biggr\} P_+.
\eea
As before, $f_{\ldots}$, $g_{\ldots}$ and $h_{\ldots}$
indicate unpolarized, chirality and transverse spin distributions.
The subscripts $L$ and $T$
indicate the target polarization, and the superscript $\perp$ signals 
explicit presence of transverse momentum of partons.
Using the notation 
$f^{(1)}(x,\bm p_\st^2) \equiv (\bm p_\st^2/2M^2)\,f(x,\bm p_\st^2)$,
one sees that $f_1(x,\bm p_\st^2)$, $g_1(x,\bm p_\st^2) 
= g_{1L}(x,\bm p_\st^2)$ and $h_1(x,\bm p_\st^2) 
= h_{1T}(x,\bm p_\st^2) + h_{1T}^{\perp (1)}(x,\bm p_\st^2)$
are the functions surviving $p_\st$-integration.

Analogously, $\Delta$ is parametrized in terms of 
unpolarized, chirality and transverse-spin fragmentation 
functions~\cite{MT96},
denoted by capital letters $D_{\ldots}$, $G_{\ldots}$, and $H_{\ldots}$, 
respectively.
Time-reversal invariance has not been imposed in the above parametrization,
allowing for non-vanishing T-odd functions $f_{1T}^\perp$ and $h_1^\perp$. 
Possible sources of T-odd effects in the initial state have been 
discussed in Refs~\cite{Sivers90}. In the final state time-reversal
invariance cannot be imposed~\cite{RKR71,HHK83,JJ93}, 
leading to non-vanishing fragmentation functions
$D_{1T}^{\perp}$~\cite{MT96} and $H_1^\perp$~\cite{Collins93}. 

To put bounds on the transverse momentum dependent functions, we
again make the matrix structure explicit. One finds for 
$M = (\Phi(x,p_\st)\,\gamma^+)^T$
the full spin matrix $\tilde M$ to be
\[
\left\lgroup \begin{array}{cccc}
f_1 + g_{1L} &
\frac{\vert p_\st\vert}{M}\,e^{i\phi}\left(g_{1T}+i\,f_{1T}^\perp\right)&
\frac{\vert p_\st\vert}{M}\,e^{-i\phi}\left(h_{1L}^\perp+i\,h_1^\perp\right)&
2\,h_1\\
& & & \\
\frac{\vert p_\st\vert}{M}\,e^{-i\phi}\left(g_{1T}-i\,f_{1T}^\perp\right)&
f_1 - g_{1L} &
\frac{\vert p_\st\vert^2}{M^2}e^{-2i\phi}\,h_{1T}^\perp &
-\frac{\vert p_\st\vert}{M}\,e^{-i\phi}\left(h_{1L}^\perp-i\,h_1^\perp\right)\\
& & & \\
\frac{\vert p_\st\vert}{M}\,e^{i\phi}\left(h_{1L}^\perp-i\,h_1^\perp\right)&
\frac{\vert p_\st\vert^2}{M^2}e^{2i\phi}\,h_{1T}^\perp &
f_1 - g_{1L} &
-\frac{\vert p_\st\vert}{M}\,e^{i\phi}\left(g_{1T}-i\,f_{1T}^\perp\right)\\
& & & \\
2\,h_1 &
-\frac{\vert p_\st\vert}{M}\,e^{i\phi}\left(h_{1L}^\perp+i\,h_1^\perp\right)&
-\frac{\vert p_\st\vert}{M}\,e^{-i\phi}\left(g_{1T}+i\,f_{1T}^\perp\right) &
f_1 + g_{1L}
\end{array}\right\rgroup ,
\]
where $\phi$ is the azimuthal angle of $\bm p_\st$.
First of all, this matrix is illustrative as it shows the full quark 
helicity structure accessible in a polarized nucleon~\cite{BoM99}, 
which is equivalent to the full helicity structure of the forward 
antiquark-nucleon scattering amplitude.
Bounds to assure positivity of any matrix element can for instance be
obtained by looking at the 1-dimensional subspaces, giving the
the trivial bounds $f_1 \ge 0$ and $\vert g_{1L}\vert \le f_1$.
From the 2-dimensional subspace one finds
(omitting the $(x,\bm p_\st^2)$ dependences)
\bea
&& \vert h_1 \vert \le
\frac{1}{2}\left( f_1 + g_{1L}\right)
\le f_1,
\\
&&
\vert h_{1T}^{\perp(1)}\vert \le
\frac{1}{2}\left( f_1 - g_{1L}\right)
\le f_1,
\\
&& \left( g_{1T}^{(1)}\right)^2
+ \left( f_{1T}^{\perp (1)}\right)^2
\le \frac{\bm p_\st^2}{4M^2}
\left( f_1 + g_{1L}\right)
\left( f_1 - g_{1L}\right)
\le \frac{\bm p_\st^2}{4M^2}\,f_1^2,
\\
&& \left( h_{1L}^{\perp (1)}\right)^2
+ \left( h_{1}^{\perp (1)}\right)^2
\le \frac{\bm p_\st^2}{4M^2}
\left( f_1 + g_{1L}\right)
\left( f_1 - g_{1L}\right)
\le \frac{\bm p_\st^2}{4M^2}\,f_1^2.
\eea
Besides the Soffer bound, new bounds
for the distribution functions are found. In particular, one sees that 
functions like
$g_{1T}^{(1)}$ and $h_{1L}^{\perp (1)}$ appearing in azimuthal asymmetries
in leptoproduction are proportional to $\vert \bm p_\st\vert$ for small
$p_\st$. In the case of the T-odd fragmentation functions, 
the Collins function,
$H_1^{\perp (1)}$, describing fragmentation of a transversely 
polarized quark into an unpolarized or spinless hadron, for 
instance a pion, is bounded by 
$(\vert \bm P_{\pi\perp}\vert/2zM_\pi) D_1(z,\bm P_{\pi\perp}^2)$ 
while the other T-odd function $D_{1T}^{\perp (1)}$ 
describing fragmentation of an unpolarized quark into a polarized 
hadron such as a $\Lambda$, is given by 
$(\vert \bm P_{\Lambda\perp}\vert/2zM_\Lambda) 
D_1(z,\bm P_{\Lambda\perp}^2)$.

Before sharpening these bounds via eigenvalues, it is convenient to
introduce two positive definite functions $F(x,\bm p_\st^2)$ and 
$G(x,\bm p_\st^2)$ such that $f_1 = F + G$ and $g_1 = F - G$ and define
\bea
&&
h_1 = \alpha\,F ,
\\ &&
h_{1T}^{\perp (1)} = \beta\,G ,
\\ &&
g_{1T}^{(1)} + i\,f_{1T}^{\perp (1)} =
\gamma\,\frac{\vert p_\st\vert}{M}\,\sqrt{FG} ,
\\ &&
h_{1L}^{\perp (1)} + i\,h_{1}^{\perp (1)} =
\delta\,\frac{\vert p_\st\vert}{M}\,\sqrt{FG} ,
\eea
where the $x$ and $\bm p_\st^2$ dependent functions $\alpha$, $\beta$, 
$\gamma$ and $\delta$ 
have absolute values in the interval $[-1,1]$. Note that
$\alpha$ and $\beta$ are real-valued but
$\gamma$ and $\delta$ are complex-valued, the imaginary part
determining the strength of the T-odd functions. Actually, one sees
that the T-odd functions $f_{1T}^{\perp}$ and $h_1^\perp$ could be
considered as imaginary parts of $g_{1T}$ and $h_{1L}^{\perp}$,
respectively.

\begin{figure}[t]
\begin{center}
\epsfig{file=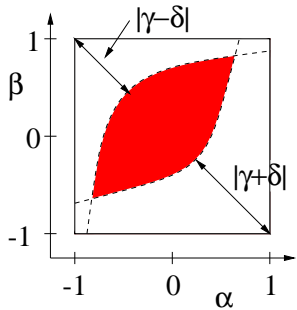,width = 4.0 cm}
\end{center}
\caption{\label{fig3}
Allowed region (shaded) for $\alpha$ and $\beta$ depending on
$\gamma$ and $\delta$.}
\end{figure}

Next we sharpen these bounds using the eigenvalues of the
matrix, which are given by
\bea
e_{1,2} = (1-\alpha)F + (1+\beta)G
\pm \sqrt{4FG\vert\gamma+\delta\vert^2+((1-\alpha)F-(1+\beta)G)^2},
\\
e_{3,4} = (1+\alpha)F + (1-\beta)G
\pm \sqrt{4FG\vert\gamma-\delta\vert^2+((1+\alpha)F-(1-\beta)G)^2}.
\eea
Requiring them to be positive can be converted into the conditions
\bea
&&
F+G\ge 0.
\\ &&
\vert \alpha\,F-\beta\,G \vert \le F+G,
\quad \mbox{i.e.} \ \vert h_{1T}\vert \le f_1
\\ &&
\vert \gamma + \delta \vert^2 \le (1-\alpha)(1+\beta) ,
\\ &&
\vert \gamma - \delta \vert^2 \le (1+\alpha)(1-\beta) .
\eea
It is interesting for the phenomenology of deep inelastic processes that 
a bound for the transverse spin distribution $h_1$ is provided not 
only by the inclusively measured functions $f_1$ and $g_1$, but also
by the functions $g_{1T}$ and $h_{1L}^{\perp}$, responsible
for specific azimuthal asymmetries~\cite{MT96,BM98}.
This is illustrated in Fig.~\ref{fig3}. 
The same goes for fragmentation functions, where for instance the 
magnitude of $H_1^\perp$ constrains the magnitude of 
$H_1$~\cite{J96}.
Recently SMC~\cite{SMC}, HERMES~\cite{HERMES} and LEP~\cite{LEP} 
have reported preliminary results for azimuthal asymmetries. More results
are likely to come in the next few years from HERMES, HERA, RHIC and 
COMPASS experiments. Although much theoretical work is needed, 
for instance on factorization, scheme ambiguities and 
the stability of the bounds under evolution~\cite{scheme}, 
these future experiments may provide us with the knowledge of the
full helicity structure of quarks in a nucleon. 
The elementary bounds derived in this paper can serve as 
important guidance to estimate the magnitudes of 
asymmetries expected in the various processes.

\acknowledgments
We would like to thank Elliot Leader for useful discussions.
This work is part of the research program of the
Foundation for Fundamental Research on Matter (FOM)
and the Dutch Organization for Scientific Research (NWO).
It is also part of the TMR program ERB FMRX-CT96-0008.


\begin{thebibliography}{30}
\bibitem{remark}
Other common notations for a quark flavor $q$ are $f^q_1(x) = q(x)$, 
$g^q_1(x) = \Delta q(x)$ and $h^q_1(x) = \Delta_\st q(x)$.
\end{thebibliography}
\end{document}